\documentclass[12pt]{article}
\usepackage{hyperref}

\usepackage{graphicx}
\begin{document}

\title { Optical potentials of halo and weakly bound nuclei}

\author { A. Bonaccorso$^{(a)}$ and F. Carstoiu$^{(b)}$\\
\small $^{(a)}$ Istituto Nazionale di Fisica Nucleare, Sezione di
Pisa, 56100 Pisa, Italy.\\
\small  $^{(a)}$ Institute of Atomic Physics, P. O. Box MG-6, Bucharest,
Romania.}

\maketitle

\begin{abstract}

The optical potential of halo and weakly bound nuclei has a long range
part due to the coupling to breakup  that damps the   elastic
scattering angular distributions at all angles for which the effect of the nuclear
interaction is felt. In charge exchange reactions leading to
 a final state with a halo
nucleus, the surface potential is responsible for a strong reduction in
the absolute cross section. We show how the halo effect can be simply estimated
semiclassically and related to the properties of the halo wave function. Assuming an exponential
tail for the imaginary surface potential  we show that the most important
parameter is the diffusness
$\alpha$ of the potential which is directly related to the decay length
$\gamma_i$ of the initial wave function by $\alpha\approx(2\gamma_i)^{-1}$.
 
\end{abstract}
\vspace{1em}
\section{Introduction}

 In the last  years
since the advent of Radioactive Beams (RIBs) \cite{ta} a new phenomenon
called 'nuclear halo' \cite{hjj} has appeared in nuclear physics.  In  typical halo
nuclei such  as $^{11}$Be, $^{19}$C or $^{8}B$ \cite{ws}-\cite{vm1}
 the valence neutron (proton) or the last couple of neutrons, as in $^{11}$Li, occupy 
weakly bound  single particle states of low angular momentum (s
or p). 
The single particle wave function of a nucleon halo has a long tail
which extends mostly outside the potential well. Then 
 the reactions initiated by such nuclei give large
reaction cross sections and  neutron breakup cross sections. Also the
ejectile parallel momentum distributions following breakup can be very
narrow, typically $40-45MeV/c$.

Elastic nucleus-nucleus scattering with a radioactive projectile \cite{exp} is
another reaction which has been studied to some extent in the attempt to
find characteristics that would be typical for a weakly bound nucleus
and would help understanding the halo structure. It has been established that
the halo breakup is responsible for a damping in the elastic angular distribution in the
 range 5$^o$-20$^o$ about.
Recently charge exchange reactions  which produce radioactive
nuclei in the final state, have also been studied. The effect of the halo breakup is very
dramatic in this case, reducing the absolute cross sections by about 50\%\cite{cap}.

All theoretical methods used to describe the above mentioned reactions,
require at some stage of the calculation the knowledge of the
nucleus-nucleus optical potential. The optical potential is the
basic ingredient for the description of elastic scattering, but it is
important also in breakup calculations, since we need to take into
account the core quasi-elastic scattering by the target while the halo
neutron breaks up. Furthermore in some breakup reactions like those
initiated from $^{12}$Be or from a core orbital of $^{11}$Be it is the
ejectile that most likely is going to have a halo structure \cite{jat}. In the charge
exchange  reaction
$^{11}B(^{7}Li,^{7}Be)^{11}Be$  the halo nucleus-nucleus optical potential
necessary to describe the final channel\cite{cap} has a volume part obtained
with a double folding plus a very diffuse surface term fitted
phenomenologically to reproduce the final channel angular distribution.

The problem of the determination of the optical potential for a halo
projectile has already been studied by many authors and a review of the present situation can be found
in \cite{pv}. One method is to start  from a phenomenologically determined core-target
potential and then the effect of the breakup of the halo neutron is added. This
process leads to adding a surface part to the core-target potential.
This new  surface peaked optical potential has been seen to have a
quite long range which should reflect the properties of the long tail
of the halo neutron wave function. Such kind of potentials are often called
dynamical polarization potentials.

 The papers published so far can be
divided in two categories: those in which the potential is calculated
microscopically \cite{Tak}-\cite{jat3}, and those in which it is obtained
phenomenologically by fitting  elastic or quasielastic data  \cite{pv,huss,so}.

In this contribution we propose a new approach to the calculation of
the imaginary part of the optical potential due to breakup. It is based
on  a  semiclassical  method described  by Broglia and
Winter  in \cite{bwb,bpw} and used also by Brink and collaborators \cite{bpb,sb} to calculate
the surface optical potential due to transfer and on the Bonaccorso and
Brink  model for transfer to the continuum reactions\cite{bb}-\cite{abb}. The
latter is based on the idea that breakup is a reaction following the same
dynamics as transfer but leading manly to continuum final states for incident
energies per nucleon higher that the average nucleon binding energy.  The
calculations are almost completely analytical and we will show that a simple
approximated formula can be obtained which will help us discussing the origin
of the long range nature of the potential and its dependence on the incident
energy as well as on the initial neutron binding energy. The characteristics of
our potential are consistent with those of   potentials obtained with other
methods, in particular our theory is close in spirit to the eikonal method of
Canto et al.\cite{cd} and application to the description of experimental data
are encouraging.
\section {Theory} 

The method we will present here is based on the extraction of an optical potential
from the calculation of a phase shift.

 The elastic scattering probability is
$P_{el}=|S_{NN}|^2$,  given in terms of the nucleus-nucleus S-matrix.
We know that
 \begin{equation}|S_{NN}(b)|^2=e^{-4 \delta_I(b)}.\label{a}\end{equation}
In a semiclassical approximation \cite{bwb}, the imaginary part of the nucleus-nucleus
phase shift
$\delta_I$ is related to the imaginary part of the optical potential by 

\begin{equation}\delta_I(b)=-{1\over 2\hbar}
\int_{-\infty}^{+\infty}\left(W_V({\bf r}(t))+W_S({\bf r}(t))\right)dt
\label{b}\end{equation}
where the volume potential is responsible for the usual inelastic core-target
interaction, while the surface term takes care of the peripheral reactions like
transfer and breakup.
${\bf r}(t)={\bf b}+vt$ is the classical trajectory of relative motion for the
nucleus-nucleus collision.

According to \cite{bwb}-\cite{sb} the surface optical potential
$W_S({\bf r}(t))$ due to
transfer can be related to the transfer probability by 
\begin{equation}
\int_{-\infty}^{+\infty}W_S({\bf r}(t))dt=-{\hbar\over
2}\sum_{(i,n)}P_n^{(i)}\label{1}
\end{equation}
where  $P_n^{(i)}$ are the transfer probabilities in
the various channels n. In the traditional formulation the index (i)
stands for stripping and pickup to bound states, here we extend it to
hold for breakup reactions in which the final neutron state is in the
continuum. Breakup of both absorptive and diffractive type will be
included. Absorptive breakup has often been called stripping within
the halo community. The justification of the use of Eq.(\ref{1}) to calculate the imaginary
potential due to breakup is simply given by the analogy between breakup and transfer as 
expressed by
 the
transfer to the continuum model introduced in Refs.\cite{bb}-\cite{abb}. There it was shown that the
formalism for transfer to bound states goes over transfer to the continuum in a natural way if the
kinematics of the reaction is taken into account correctly within a time dependent approach which
ensures neutron energy conservation.

Using  Eq.(\ref{b}) and (\ref{1}), in (\ref{a}) the
nucleus-nucleus S-matrix, in the case of a halo projectile, can be written as
 
\begin{equation}|S_{NN}|^2=|S_{CT}|^2 e^{-P_{b_{up}}}\label{s}\end{equation}
where $S_{CT}$ takes into account all core-target interactions while the term
$e^{-P_{b_{up}}}$ depends only on the halo neutron breakup probability. 
For a halo nucleus at high incident  energy the transfer
probability is going to be much smaller than the breakup probability, therefore the surface
potential has be identified here with the breakup potential. 

In order to obtain the surface imaginary potential equation (\ref{1}) should be calculated
as an identity in the distance of closest approach, which amounts to require that $ W_S(r)$
be a local, angular momentum independent function. We remind the reader that  since
we are using a semiclassical method, the non locality, which is in
principle a characteristic of microscopic optical potentials
 has been transformed into an energy dependence\cite{bbopt}.

Now we discuss the hypothesis leading to Eq.(\ref{s}). The reactions with halo projectiles we are concerned with in this
paper have been performed at  energies well above the Coulomb barrier where many inelastic channels open at about the same
distance of closest approach.
The effect of the breakup is most important at large distances ($b>R_s$) of
closest approach,
 where it represents the dominant reaction mechanism. If the breakup probability is
needed at smaller impact parameters, then the values  calculated by
perturbation theory, have to be multiplied by the core survival probability, as discussed
in Eq.(V.8.1) of Broglia and Winther and also used in relation to halo breakup by several
authors. 
 The effect of all inelastic channels $n$ different
from the one we are interested in, can be taken into account by
introducing a damping factor $P_0$. Therefore the breakup probability
$P_{b_{up}}$ at all distances can be defined as

\begin{equation} P_{b_{up}}=p_{b_{up}}\prod_n(1-p_n)\approx p_{b_{up}}
\exp(-\sum_n p_n)=p_{b_{up}}P_0\label{2}\end{equation}
Each elementary inelastic probability $p_n$ and breakup probability
$p_{b_{up}}$ is  small and $p_{b_{up}}$ in particular, can be calculated in
time dependent perturbation theory, as done in
\cite{bb}. In this paper we will treat only the nuclear breakup channels, which
are important for light targets. In the case of heavy targets also the
Coulomb breakup has to be taken into account.

 In reactions with halo projectiles the damping factor $P_0$ has also
been referred to as the core survival probability 
after the halo breakup or as the core elastic scattering probability. The 
  breakup probability Eq.(\ref{2}) integrated over the impact parameter b  has been
widely used in the literature to get total breakup cross sections.

The breakup probability $p_{b_{up}}$ with the index $b_{up}$ standing for one
neutron breakup can be obtained by integrating the neutron energy or
momentum spectrum as given for example in \cite{abb}.

\begin{equation}{p_{b_{up}}}
\approx \int d\varepsilon_f\Sigma_{l_f}(|1-\langle S_{l_f}\rangle
|^2+1-|\langle S_{l_f}\rangle |^2) B(l_f,l_i). 
\label{dpde}\end{equation}

It is important to remark that the above expression takes into account to all orders the neutron target
final state interaction via an energy and angular momentum dependent optical model wave function of the
breakup neutron. In this way neutron elastic scattering and absorption are treated
consistently via an unitary S-matrix. 
Eq.(\ref{dpde}) is  the neutron transfer probability from a definite single
particle state of energy
 $\varepsilon_i$, momentum $\gamma_i=\sqrt
{-2m\varepsilon_i}/\hbar$, and angular momentum $l_i$  in the
projectile to all possible final continuum state of energy
$\varepsilon_f$, momentum $k_f=\sqrt {2m\varepsilon_f}/\hbar$. It is
the sum of the transfer probabilities to each possible final
$l_f$-state for a given  final energy  $\varepsilon_f$. In
Ref.\cite {bb1}  it was shown that the first term of
 Eq.(\ref{dpde}), proportional to
 $|1-\langle S_{l_f}\rangle |^2$ , gives the neutron elastic breakup
spectrum while the second term proportional to the transmission
coefficient $T=1-|\langle S_{l_f}\rangle |^2$ gives the absorption
spectrum. 
 This term contains contributions from inelastic scattering of the
breakup neutron by the target nucleus and also from compound nucleus
formation.

The factor $B(l_f,l_i)$ is an elementary transfer
probability which depends on the details of the initial and final
states, on the energy of relative motion and on the distance of
closest approach $b$ between the two nuclei. Its explicit expression
reads:

\begin{equation}B_{l_f,l_i}={1\over 2}\left [{\hbar\over mv}\right ]^2{m\over
\hbar^2k_f}(2l_f+1)|C_i|^2 {e^{-2\eta b}\over 2\eta b}P_{l_i}(X_i)P_{l_f}(X_f),
\label{B}\end{equation}
where $X_i=2(\eta/\gamma_i)^2-1$,
 $X_f=2(\eta/k_f)^2+1$. Also
$k_1=-(\varepsilon_i-\varepsilon_f+{1\over 2}mv^2)/(\hbar v)$ and
$k_2=-(\varepsilon_i-\varepsilon_f-{1\over 2}mv^2)/(\hbar v)$ are the
 $z$ components of the neutron momentum in the initial and final state,
respectively. $\eta^2=k_1^2+\gamma_i^2=k_2^2-k_f^2$ is the modulus square of the
transverse component of the neutron momentum. $mv^2/2$ is the incident energy 
per
nucleon at the distance of closest approach $b$ for the ion-ion collision. 
$|C_i|^2$
is the asymptotic normalization constant of the initial state wave function 
 and $P_{l_i}$ and $P_{l_f}$ are Legendre polynomials coming from the 
angular parts
of the initial and final wave functions respectively \cite{bb}.
 Coulomb breakup can  be taken into account as well, 
following the formalism of
\cite{jer}. One advantage of calculating the breakup probability by Eq.(\ref{B})  is that no sudden
approximation hypothesis is made and thus the method is valid also for any initial separation energy.

In Eqs.(\ref{dpde}) the main dependence on the core-target
distance of closest approach $b$ is contained in the exponential factor
$e^{-2\eta b}$.  Equation (\ref{B}) has a maximum in correspondence
to the minimum value of  $\eta=\gamma_i$. Therefore after integrating
over $\varepsilon_f$  the $b$-dependence of the breakup probability
$p_{b_{up}}(b)$ will still be of the exponential form
$p_{b_{up}}(b)\approx  e^{-b/{\alpha}}$ with  $\alpha\approx
(2\gamma_i)^{-1}$ where $\gamma_i$ is the decay length of the neutron
initial state wave function. We now assume at large distances, where $P_0=1$
the same exponential dependence for the absorptive potential,
$W_S(r)=W_0e^{-r/{\alpha} }$ and as indicated earlier on, a straight line
parameterization for the trajectory ${\bf r}(t)={\bf b}+vt$, then 
Eq.(\ref{1})  reads
\begin{equation}
\int_{-\infty}^{+\infty}W_S( b,z)dz=-{\hbar v\over
2}p_{b_{up}}(b).\label{1bis}
\end{equation} The LHS can be approximately evaluated as 

\begin{eqnarray}\int_{-\infty}^{+\infty}
W_S(b,z)dz={W_0}\int_{-\infty}^{+\infty}e^{-(b+{z^2\over
2b})/\alpha}dz  ={W_0}\sqrt{2\pi
b\alpha}e^{-b/\alpha},\label{09}\end{eqnarray}
where we assumed $b>>z$ in the second step.
Equating the RHS of Eqs.(\ref{1bis}) and (\ref{09})  and renaming the
distance $b$ as $r$ gives

\begin{eqnarray}W_S(r)&=&
-{\hbar v\over 2}p_{b_{up}}(r){1\over \sqrt{2
\pi \alpha r}}\label{9}\end{eqnarray}

The exponential form of $W_S(r)$ implies that the strength of the breakup potential
be a function of $r$. However we know that in nuclear induced peripheral reactions
like breakup and transfer  most of the cross section comes from impact parameters
around the strong absorption radius. Therefore  writing
$p_{b_{up}}(r)\approx
p_{b_{up}}(R_s)e^{-(r-R_s)/{\alpha}}$ we finally get that Eq.(\ref{9}) can be
written as

\begin{equation}
W_S(r)\approx W_0e^{-{r-R_s\over \alpha}}\label{10} \end{equation}
 where
\begin{equation}W_0\equiv W_0(R_s)=-{\hbar v\over 2}p_{b_{up}}(R_s){1\over
\sqrt{2 \pi \alpha R_s}},\label{11}\end{equation} which gives an
estimate of the strength parameter of the surface breakup potential at the typical
distance $R_s$.

Equation (\ref{10}) has a number of interesting features. First of all it
shows explicitly that the long range nature of the breakup potential
originates from the large decay length of the initial state wave
function. For a typical halo separation energy of  0.5MeV,
$\alpha=(2\gamma_i)^{-1}=3.2fm$, while for a 'normal' binding energy of 10MeV, $\alpha=0.7fm$ as
expected. Therefore the parameter
$\alpha$ will depend only on the projectile characteristics and not on the target. Furthermore looking
at Eqs. (\ref{10}) and (\ref{11}) we notice that for a fixed initial state  the strength of the
potential will be larger the smaller the neutron binding energy. On the other hand for a fixed binding
energy the potential strength will be lower the higher the initial angular momentum. Finally  the
strength parameter
$W_0$ is seen to be energy dependent from  different sources. If we consider $R_s$, the
typical distance  at which the
strength is calculated, to be the same at all energies, then the energy dependence of  $W_0$
is given by its linear dependence on  the velocity of relative motion
$v$, which is a function  of the projectile-target combination.
Another energy
dependence   is trough the breakup  probability, whose behaviour in 
turn is
 determined in part by the neutron-target energy dependent optical potential.
At the large distances  we are interested in, the overall energy dependence of
the  breakup probability is an  exponential decrease with incident energy due
to the dependence on
$v$ and on the neutron-target optical potential. Therefore
$W_0$ is expected to rise up to about 40 A.MeV and then to decrease at higher
energies. Another interesting way to look at the behaviour of the  strength 
$W_0$ is to consider instead explicitly that the strong absorption radius $R_s$
is itself decreasing with energy. If we take this dependence  into account,
then $W_0$ increases up to about 80 A.MeV and then starts to decrease. However 
the precise energy dependence of $R_s$ requires an accurate
 knowledge of the energy dependence  of the core-target volume optical potential. This is
beyond the scope of the present work, therefore we will discuss the energy dependence of the
surface potential at a fixed distance large enough to have unit core survival probability
at all incident energies.

It is very well known that the dynamical polarization potential due to surface
reactions gives rise also to a correction to the real part of the nucleus-nucleus
optical potential. In terms of Feshbach potential both the real and imaginary parts
of the dynamical polarization potential come from the second order term and
therefore the real polarization term gives a correction to the first order term,
which is purely real and it is often referred to as the folding potential. The relative magnitude of this
correction with respect to the first order real potential depends on the system involved and on the
incident energy. One characteristic of the real dynamical potential, discussed by many authors is to become
repulsive at some energies. The imaginary polarization potential on the other hand is by definition negative
from Eq.(\ref{9}). The real dynamical potential is expected to have the same exponential dependence
as the imaginary part.  The simple and consistent way we have used here to obtain
 its strength is by applying a dispersion relation. 

  \subsection{Dispersion relation}

The theoretical optical potential is highly nonlocal and energy
dependent. In most  applications it is replaced by an equivalent local
potential $U(r,E)=V_R(r,E)+\Delta V(r,E)+iW(r,E)$. The term $V_R$ is
usually associated with the folding potential and contains a spurious
energy dependence due to the finite range of the underlying nucleon-nucleon effective
interaction and Pauli principle. We have further assumed that the imaginary
potential  splits  into contributions
coming from coupling to breakup ($W_S$) and other inelastic excitations ($W_V$) so that
$W(r,E)=W_{V}(r,E)+W_{S}(r,E)$. The real part of the dynamic polarization potential (DPP)
arising from coupling of elastic channel to the breakup is calculated
from the dispersion relation
\begin{equation}
\Delta V(r,E)=\frac{P}{\pi}\int_0^{\infty}\frac{W_{S}(r,E')dE'}{E'-E}.
\label{disp}\end{equation}
The numerical evaluation of this term requires the knowledge of the
imaginary potential at all energies. Our model provides accurate values in
a limited range of energies $E<1600$ MeV for which the nucleon target
potential is known with reasonable accuracy. At higher energies we assume a
reasonable energy dependence of the form $W_{S}\sim E^{-1}$ in such a way
that the integral in Eq. (\ref{disp}) converges and it can be evaluated accurately for
the energies of interest. An algebraic exact model similar to that used in
\cite{mah} has been used to check the numerical accuracy.

\section {Results}

In order to sample the quantitative accuracy of the simple analytical model presented
above we discuss now some numerical examples. The potentials we will discuss derive from
the breakup of the $2s_{1/2}$ and $1p_{1/2}$ states  of $^{11}Be$, with separation energies
$0.5MeV$ and $0.18MeV$ respectively. It
has been shown that during the charge exchange reaction of \cite{cap}, $^{11}Be$ can be
populated in the final channel in either the ground state or the first $1/2^-$ excited
state. There are in the literature a number of papers discussing the breakup potential
for a $^{11}Li$ projectile, among others \cite{jat3,pv,yab,cd}. $^{11}Be$
breakup from the $2s$ has been discussed in \cite{jat1,jat2}. However the
potential due to breakup of the $1p_{1/2}$ bound excited state has never been
discussed before.

The neutron-target optical potential used here to calculate the breakup
probabilities is the same as in \cite{BB1}. The core survival probability has
been parameterized as
\begin{equation}
P_0(b)=|S_{CT}|^2=exp(-ln2e^{[(R_s-b)/a]}),\label{pel}\end{equation}
where $a=0.6fm$ and the strong absorption radius $R_s=1.4
(A_P^{1/3}+A_T^{1/3})fm$.

\begin{figure}[ht]
\centering
\includegraphics[scale=0.5,angle=90]{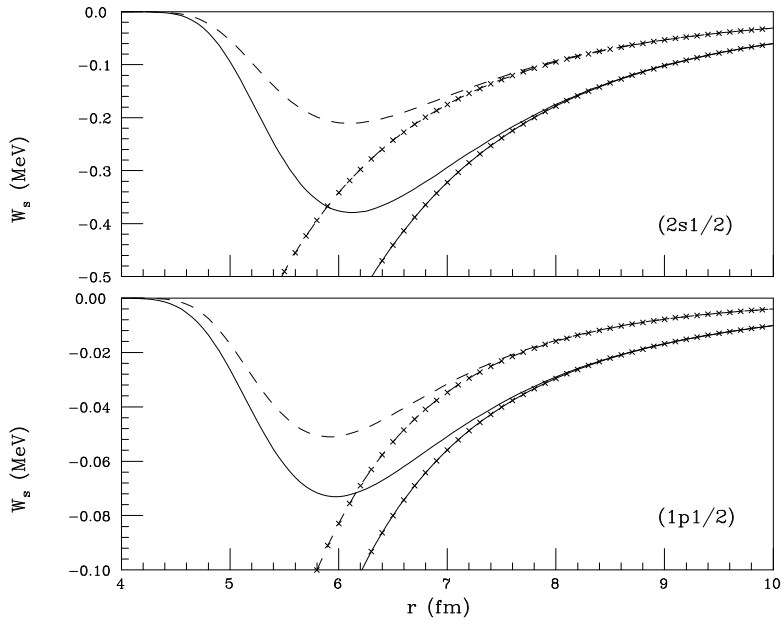}
\caption{Radial shape of the surface imaginary potential for the system $^{7}Be+^{11}Be$ due to breakup
from the $2s$ and
$1p_{1/2}$ states of $^{11}Be$ at $E=57MeV$ (dashed lines) and 
 $E=550MeV$ (solid lines). Lines with symbols are calculated assuming $P_0=1$ in Eqs.(\ref{1}) and
(\ref{2}).}
\end{figure}
We start by showing in Fig.(1) the radial shapes of the potentials calculated for the
breakup from the 
$2s$ and $1p_{1/2}$ states of $^{11}Be$ in the interaction with
$^{7}Be$ relevant to the charge exchange reaction of
\cite{cap}. In both figures we show results for two
laboratory scattering energies for the ion-ion system:  $E=57MeV$ (dashed lines) and 
 $E=550MeV$ (solid lines). The lines with symbols correspond to the exponential approximation
for the potentials, Eqs.(\ref{9}), (\ref{10}), (\ref{11}) while the lines without symbols are obtained
from Eq.(\ref{9}) using for the breakup probability the $ P_{b_{up}}\approx
p_{b_{up}}P_0$ definition valid at all distances. 
The potential due to the $1p_{1/2}$ state breakup is about one order of
magnitude weaker than the potential due to the $2s$ state, although the binding
energy is smaller. This is due to the fact that for $l>0$ states the effect of
the centrifugal barrier hinders breakup.  Our results show  a strong dependence of
the potential on the incident energy   (compare solid and
dashed lines) and also a quite strong state  dependence. In fact  in the case of the
s-state the diffusness of the potential is about 2.3fm, while for the p-state is about 2fm
while  the strength for  the s-state is about five times more than for the p-state.
The values of the diffusness are slightly smaller than the estimate given in Sec.2 because
of the integration over the neutron  final energy of the breakup probability. Finally we
would like to stress that the internal part of the surface polarization potential has no
effect on the nucleus-nucleus  scattering as it can be seen also from the nucleus-nucleus
S-matrix  of Fig. (2) which we are going to discuss next.
\begin{figure}[ht]
\centering
\includegraphics[scale=0.35,angle=90]{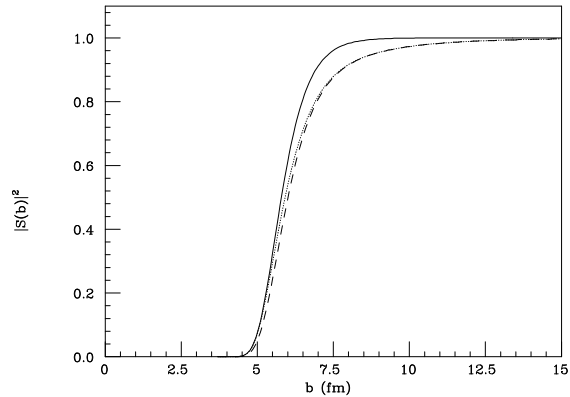}
\caption{S-matrix values as a function of the impact parameter for the system 
$^{11}Be+^{9}Be$ at 50A.MeV. Solid line is $|S_{CT}|^2$, dashed and dotted lines are $|S_{NN}|^2 $
calculated with the two prescriptions for the breakup probability discussed in the text. }
\end{figure}

Another important application is in fact to see how the elastic scattering total probability
changes as a function of the impact parameter or angular momentum when there is a strong
breakup probability.  In Fig.(2) we show the core-target S-matrix, $S_{CT}$ of Eq.(\ref{pel})
(solid line)  and the nucleus-nucleus
S-matrix
$S_{NN}$ (dashed line) from Eq.(\ref{s}), calculated with $P_{b_{up}}= p_{b_{up}}$, which contains
the effect of the halo breakup. In this case the exponential approximation for the surface
potential is used at all distances. At a fixed impact parameter ( or angular momentum) the effect of
the breakup is to reduce the elastic probability given by the modulus square of the S-matrix.  The unitarity limit is attained
at much larger b-values and the reaction cross section receives
significant contributions from a large range of impact parameters. This
result is analogous to the discussion reported in
\cite{jat}. The reduction is more pronounced at the impact parameters larger
than the strong absorption radius.    
  The value of the strong absorption radius does not change appreciably because
it is mainly determined by the core-target interaction which is strongly absorptive. On the other
hand it is in a sense obvious that surface reactions such as breakup would change the S-matrix
behaviour at  large impact parameters where they represent the dominant reaction
channels. 

The dotted line is the S-matrix also calculated from
Eq.(\ref{s}) but this time we have used $P_{b_{up}}= p_{b_{up}}P_0$ with $P_0$ given by
Eq.(\ref{pel}). With this prescription one gets an imaginary breakup potential
valid at all distances. The fact that the two S-matrices (dashed and dotted
line) are hardly  distinguishable is a proof of the fact that elastic
scattering is not sensitive to the internal part  of the polarization
potential. 
\begin{figure}[ht]
\centering
\includegraphics[scale=0.35,angle=0]{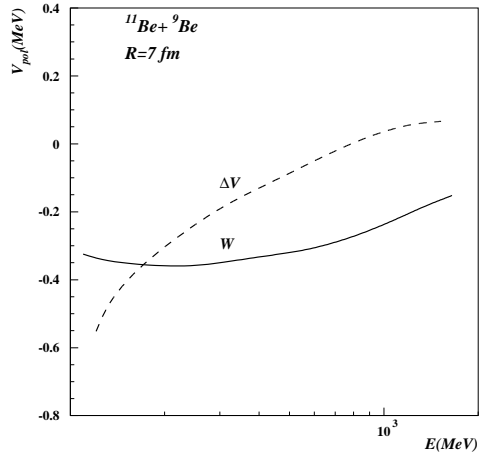}
\caption{Energy dependence of the strengths of real and imaginary polarization potential for the
system
$^{11}Be+^{9}Be$ at the distance 7 fm.  }
\end{figure}

Fig. (3) contains the energy dependence of the imaginary and real strengths of
the dynamical polarization potential due to breakup of the halo neutron in the reaction
$^{11}Be+^{9}Be$.  
 The solid line gives the energy dependence of the imaginary potential
 calculated at the fixed  distance 7 fm which
is slightly larger than the sum of the projectile and target radii. This distance is
about the smallest at which absorption into channels other than breakup can be neglected
and the core survival probability is $P_0=1$.  The real part of the potential obtained from the
dispersion relation is given by the dashed line. It shows a change of
sign which gives a repulsive real potential from around the energy (70A.MeV) at which the imaginary
part starts to have a clear decrease toward zero. This is the obvious and consistent result of having
applied the dispersion relation. Physically it means that the mean field will  prevent from entering
the interaction region those waves that cannot be absorbed.

\begin{figure}[ht]
\centering
\includegraphics[scale=0.5,angle=0]{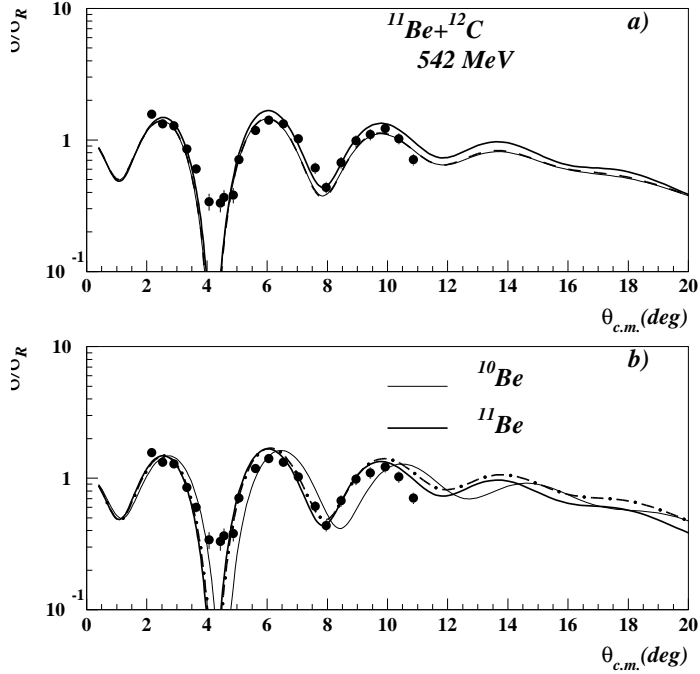}
\caption{a)Elastic scattering angular distribution for the reaction $^{11}Be+^{12}C$. Solid line is
with a bare volume imaginary potential. Dashed line is  obtained adding  the imaginary surface
potential calculated in this work. Dotted line includes also the real part of the surface potential. b)
Solid line is obtained with the same bare potential as in a),  the
dotted line is obtained by decreasing both the real and imaginary potential  radii
as explained in the text, while the dotdashed line is obtained decreasing only the imaginary potential
radius.}
\end{figure}

 Since the optical potential has
one of its most interesting application in the calculation of elastic
scattering angular distributions, we finally  show in Fig.(4a) an example for the
reaction
$^{11}Be+^{12}C$ at 49.3 A.MeV. The  data are from  P.Russel-Chomaz et al. 
\cite{prc}. The  optical
model parameters for the volume parts of the bare potential are taken  from \cite{jat2} and were fitted
to
$^{10}Be$ elastic scattering data at 59.4 A.MeV. They are

$$ V_R=123 MeV, r_R=0.75fm, a_R=0.8fm,$$

$$W_V=65  MeV, r_I=0.78fm, a_I=0.8fm.$$

 In our case we have defined the real
and imaginary radii by multiplying the radius parameters  by $(11^{1/3}+12^{1/3})$, in order to take
into account in the volume potential the presence of the extra neutron with respect to the core. The
consequences of this
 choice are discussed in the following in relation to Fig.(4b).

In Fig (4a) the solid line is the calculation 
with the bare volume  potential. The dashed line is obtained instead including  the
surface imaginary potential calculated according to the method proposed in  this work. The large
diffusivity in the breakup absorption leads to   changes in the S-matrix in
all partial waves as discussed above and the angular distribution is damped. 
The inclusion of the real
polarization potential (dotted line) gives a negligible modification to the quality of the fit since its strength
(-0.15MeV) at this incident energy (50 A.MeV) is very small with respect to the volume part. Also
 variations in the strength of the imaginary  surface potential up to about 30\% result in negligible
differences in the angular distribution.
 As expected
the effect of breakup is to suppress  scattering at all angles larger than about 5$^o$. The angular
distribution shows the usual Fraunhofer oscillations at small angles followed by an almost exponential
decrease of the cross section due to a far side dominance. No Airy like oscillation are seen since the
absorption is already too strong.

In order to clarify the dependence of the calculated angular distribution on the choice of the radius
parameters of the bare potential, in Fig (4b) we show again the data and the angular distribution with
the bare potential as in  Fig(4a) (full line), plus the angular distribution with the bare potential
in which the radii have been calculated from the above  radius parameters but multiplied by
$(10^{1/3}+12^{1/3})$ (dotted line). This is to show that, as expected, a small decrease in the radius
of the optical potential would give a slight shift toward larger angles.  With the dotdashed line we
show instead the calculation done with the radii chosen as $R_R=r_R(11^{1/3}+12^{1/3})$fm and
$R_I=r_I(10^{1/3}+12^{1/3})$fm. This calculation agrees very well with the full line calculation up to
about 10$^o$. For larger angles only a change in the magnitude of the cross section is seen while there
is no shift in the peak position, which is then determined by the radius of the real volume potential. 

Another significant
effect of the imaginary surface potential
is seen in the calculated total reaction cross section given in  Table I. We
obtain an increase of 150 mb with respect to the bare (no breakup) optical
potential mainly due to an increase of about 10\% in the rms radius 
and 5\% in the volume integral of the absorption. The increase in the reaction
cross section is very close to the total breakup cross section
$\sigma_{b_{up}}\approx 170$ mb expected at this energy\cite{abfc}. This is
consistent with the  hypothesis that $P_{b_{up}}$ is small in Eq.(\ref{s}). In
fact expanding the exponential in Eq.(\ref{s}) to first order in $P_{b_{up}}$
and integrating over the impact parameter b one immediately finds

\begin{equation}\sigma_{NN}\approx\sigma_{CT}+\sigma_{b_{up}}\end{equation}
\begin{table}[tbp]
\caption{ Volume integrals per number of interacting nucleon
pairs and rms  radii of the Woods-Saxon potential used in Fig. 2
for  $^{11}$Be+$^{12}$C scattering. The last column gives the total
reaction cross section. Here $V_{opt}=V_R+iW_V $}
\label{tab_B}
\setlength{\oddsidemargin}{-0.5cm}
\setlength{\textwidth}{16cm}\small
\begin {center}
\vskip.7cm {\footnotesize
\begin{tabular}{||c|c|c|c|c|c||}
\hline
  &  &  &  &  &    \\
Pot. & J$_{V_R}$        & R$_{V_R}$ &
J$_{W}$        &R$_{W}$&
$\sigma_{NN}$\\
     & [MeV fm$^{3}$] & [fm]    & [MeV fm$^{3}$] & [fm]  &[mb]         \\
  &  &  &  &  &    \\ \hline
  &  &  &  &  &    \\
 V$_{opt}$                &235 &3.964 &145.6 &4.257 &1255\\
V$_{opt}+iW_{S}$          &235 &3.964 &151.7 &4.598 &1399\\
  &  &  &  &  &    \\ \hline
\end{tabular}} \end{center}\end{table}
\bigskip

In the case of the charge exchange reaction the effect of the surface breakup potential
 is more
dramatic, giving a decrease in the cross section of about 50\%\cite{cap} necessary to
fit the data.

\section{Conclusions}
 In conclusion we have presented a simple analytical method to obtain
the surface component of the real and imaginary parts of the nucleus-nucleus optical potential in the
case in which one partner of the reaction is a halo or weakly bound nucleus. The main purpose here
was to relate the characteristics of the potential to the special properties of the breakup
channel for weakly bound nuclei. The evaluation of the potential amounts in fact just
to the calculation of the breakup probability. If breakup from core excited
states is to be included, then it suffices to sum up the relative probabilities
according to Eq.(\ref{1}).

The method is an extension of that previously used to calculate microscopically the
effect of transfer channels on the imaginary potential. The shape of the
surface imaginary  potential and its
parameters are determined univocally by the shape of the breakup form factor.  An
interesting result is that the diffusness of the potential reflects the decay length of the
neutron wave function entering breakup and therefore depends mainly upon the projectile
characteristics, but in a model independent way. The strength parameter has a
rather complicated but physically understandable energy dependence which we have
discussed. At a given distance the uncertainty on the strength would be of about
30\%, reflecting mainly the model dependence of the breakup
probability values \cite{BB1}. The real potential has been obtained via the use of a
dispersion relation and shows the interesting property of becoming repulsive
when the imaginary part starts to decrease due to the closing up of the breakup
channel when the energy becomes too high. Sample calculations have shown that
the potential proposed here is consistent   with other theoretical models
Refs.\cite{yab,cd,jat2} for similar, light halo systems and also  with existing
experimental data.
  Furthermore we have
given an explicit justification for the long range of the polarization
potential as coming from the small decay length  of the initial
neutron wave function.

\end{document}